2026

# Systematic Mapping Protocol

## GENERATIVE AI-AUGMENTED REQUIREMENTS ENGINEERING: A SYSTEMATIC MAPPING STUDY OF PRACTICES, SPECIFICATIONS, AND REUSE IN SOFTWARE DEVELOPMENT

Final version: November 17th, 2026

Fernando Pinciroli[1,2]

[1]Facultad de Informática y Diseño
Universidad Champagnat, Mendoza, Argentina.

[2]Instituto de Informática
Facultad de Ciencias Exactas, Físicas y Naturales
Universidad Nacional de San Juan, Argentina.

**Abstract.** Generative artificial intelligence (GenAI), particularly large language models (LLMs), is rapidly being adopted in requirements engineering (RE). Existing secondary studies have characterized which RE activities are supported and which prompting strategies are used, but little is known about the specifications that GenAI-augmented RE produces and how they are used in subsequent activities such as specification-driven development (SDD), design and code generation, test derivation, verification, and reuse. This document presents the protocol of a systematic mapping study (SMS) that will identify and classify research on GenAI-augmented RE, with particular attention to the work products it generates and their downstream use. The protocol defines the research questions and their classification schemes, the search strategy, the study selection procedure, the data extraction form, the analysis approach, the tool support, and the threats to validity. It is published in advance to establish a prior commitment and to support the transparency and repeatability of the study.

**Keywords:** systematic mapping study; requirements engineering; generative artificial intelligence; large language models; specification-driven development; requirements reuse.

# Contents

# 1. Introduction

Requirements engineering (RE) is widely recognized as one of the most critical phases of software development, since defects introduced in requirements propagate to every subsequent activity and become increasingly costly to correct [1] [2]. RE comprises the elicitation, analysis, negotiation, specification, validation, and management of requirements, and its main work products, notably specifications, constitute the basis on which systems are designed, built, and tested [2] [3] [4].

Since the emergence of large language models capable of generating text from a small number of examples [5], generative artificial intelligence (GenAI) has been increasingly applied across software engineering [6]. RE, being largely based on natural language, has been one of the areas where this adoption has been most visible, continuing a long research tradition on natural language processing for RE [7]. Several secondary studies have recently mapped or reviewed this literature [8] [9] [10] [11], showing a strong concentration on early RE activities, such as elicitation and analysis, and on prompting strategies and evaluation methods.

At the same time, industry practice has begun to promote specification-driven development (SDD), in which structured specifications are used as the primary input for AI-assisted design, code generation, and testing [12]. This shift raises a question that the existing secondary studies do not address as their main focus: to what extent does GenAI-augmented RE produce specifications that are sufficiently precise, structured, verifiable, traceable, and reusable to serve as a reliable basis for subsequent activities of the software life cycle?

This study addresses that question by means of a systematic mapping study (SMS), following the guidelines of Petersen et al. [13] [14] and Kitchenham et al. [15]. Its central premise is that RE is the core of the process under study, and that specifications must be characterized both by their relationship to the needs they express and by their use in guiding development, testing, and reuse. Accordingly, the study does not aim to describe the use of AI throughout the life cycle in general, but to map how GenAI is used to improve RE and how RE work products are exploited downstream.

The expected contributions of the study are:

- a classified map of research on GenAI-augmented RE, organized by RE activity, process organization, work product, representation, AI integration approach, downstream use, quality control mechanism, and research type;
- a classified catalog of the practices reported in the literature;
- the identification of research concentrations and gaps, particularly regarding the downstream use of specifications;
- a foundation for future systematic literature reviews and for the design of an integrated GenAI-augmented RE process.

This protocol is structured as follows. Section 2 summarizes the background of the study. Section 3 reviews related secondary studies. Section 4 describes the research method, including the research questions, search strategy, study selection, data extraction, data analysis, and tool support. Section 5 discusses threats to validity and their mitigation. Section 6 defines the scope limitations. Section 7 describes how protocol amendments are handled. Section 8 describes the supplementary material and Section 9 concludes.

# 2. Background

## 2.1. Requirements engineering

RE is the subfield of software engineering concerned with discovering, documenting, analyzing, validating, and managing the requirements of a system throughout its life cycle [1] [2] [3]. ISO/IEC/IEEE 29148 [4] defines the characteristics of well-formed requirements, both for individual requirements (necessary, appropriate, unambiguous, complete, singular, feasible, verifiable, correct, and conforming) and for sets of requirements (complete, consistent, feasible, comprehensible, and able to be validated). This study adopts these characteristics as the reference set of quality attributes, complemented by traceability.

## 2.2. Generative artificial intelligence

GenAI refers to models capable of producing new content, such as text, code, or models, from instructions expressed in natural language. Large language models, trained on extensive corpora and adaptable to new tasks through instructions with or without examples [5], are the most widespread form of GenAI in software engineering [6]. They can be integrated into RE through different approaches, including direct prompting, chain-of-thought prompting, retrieval-augmented generation, fine-tuning, and agent-based architectures [9] [16].

## 2.3. Specification-driven development

Specification-driven development (SDD), also referred to as spec-driven development, is an emerging industrial approach in which structured specifications act as the primary artifact from which AI assistants derive plans, code, and tests [12]. Because the term originated in industry and has limited academic presence, this study does not restrict itself to that label: it includes any RE approach whose work products are used as a source, constraint, or validation criterion for subsequent activities.

# 3. Related work

A preliminary search identified the following secondary studies related to the scope of this SMS (Table 1). They were used to refine the research questions and search terms, and their primary studies will contribute to the control set used to validate the search strings (Section 4.3.3).

Table 1. Related secondary studies

| Study | Type | Period and studies | Main focus |
|---|---|---|---|
| Zhao et al. (2021) [7] | SMS | 1983–2019; 404 studies | Natural language processing for RE, prior to the widespread adoption of GenAI |
| Hou et al. (2024) [6] | SLR | 2017–2024; 395 studies | LLMs across software engineering tasks, including RE |
| Cheng et al. (2025) [8] | SLR | 2019–2024; 105 studies (extended to 238 studies up to 2025 in a later version) | GenAI applications across RE phases, models and techniques used, and implementation challenges |
| Zadenoori et al. (2025) [9] | SLR | 2023–2024; 74 studies | LLMs for RE: publication trends, supported RE activities, prompting strategies, and evaluation methods |
| Hemmat et al. (2025) [10] | SR | Not stated here | Research directions for NLP and LLMs in software requirements engineering |
| Lourenço et al. [11] | SMS | 2020–2025; 51 studies | Quality evaluation of functional requirements generated by LLMs |

These studies show that GenAI-augmented RE has been characterized mainly in terms of supported activities, models, and prompting strategies. Zadenoori et al. [9] report that code, test, and traceability artifacts account for a small share of the work products addressed in the literature. None of the studies takes as its main focus the specifications produced by GenAI-augmented RE and their use in subsequent activities. This gap justifies the present SMS, whose distinctive axes are the work products of RE (RQ3) and their downstream use, including SDD and reuse (RQ5).

# 4. Research method

The study follows the guidelines for systematic mapping studies in software engineering of Petersen et al. [13] [14] and Kitchenham et al. [15], the snowballing guidelines of Wohlin [17], and the procedure applied by the authors in a previous SMS [18] [19] [20]. Figure 1 summarizes the search and selection process.

## 4.1. Objective

This study aims to identify and classify research on requirements engineering (RE) augmented by generative artificial intelligence (GenAI). The mapping covers the activities, practices, process organization, work products, notations, and forms of AI integration reported in the literature.

Particular attention is given to specifications and to their use as input for subsequent activities, namely specification-driven development (SDD), design and code generation, test derivation, verification, and reuse.

The resulting map is intended to:

- identify areas of research concentration and gaps;
- build a classified catalog of reported practices;
- inform future systematic literature reviews and the design of an integrated GenAI-augmented RE process.

The following guiding principle governs the formulation of the research questions and the application of the selection criteria:

> *Requirements engineering is the core of the process under study. Specifications are among its principal work products, and this study characterizes them both by their relationship to the needs they express and by their use in guiding software development, testing, and reuse.*

## 4.2. Research and publication questions

A set of research questions (RQ) has been designed to accomplish this goal (Table 3), which were designed after applying the PICOC approach (Table 2).

Table 2. PICOC

| Element | Definition |
|---|---|
| Population | Studies on software requirements engineering. |
| Intervention | Use of GenAI (large language models and foundation models) in RE activities. |
| Comparison | Not applicable. |
| Outcome | Reported practices, process organization, work products, representations, downstream uses, quality control mechanisms, and research types. |
| Context | Software development in academic and industrial settings. |

Each research question is associated with a classification scheme that defines the fields of the data extraction form.

Table 3. Research questions

| RQ# | Research question | Description |
|---|---|---|
| RQ1 | Which RE activities do the studies address and which practices do they report? | • Activity: elicitation, analysis, negotiation, specification, validation, management, traceability, evolution, or reuse.<br>• Practice type: traditional practice augmented by AI, adapted traditional practice, or emerging practice. |
| RQ2 | How is the RE process organized in the reported approaches? | • Structure: single task, sequential workflow, iterative workflow, or multi-agent system.<br>• Degree of automation: assistance, collaboration, supervised autonomy, or full autonomy.<br>• Point of human intervention: definition, review, approval, or correction. |
| RQ3 | Which work products are generated, transformed, or related? | • Artifact: user stories, use cases, functional and non-functional requirements, acceptance criteria, business rules, conceptual and domain models, data dictionaries, or structured specifications.<br>• Level of formalization: informal, semi-structured, structured, or formal. |

| RQ# | Research question | Description |
|---|---|---|
| RQ4 | Which notations, AI models, integration approaches, and tools are reported? | • Notation or format: unconstrained natural language, template-based natural language, UML, BPMN, SysML, formal language, or structured format.<br>• Model family: by provider and by availability (proprietary or open).<br>• Integration approach: zero-shot prompting, few-shot prompting, chain-of-thought prompting, retrieval-augmented generation, fine-tuning, or agent-based approaches.<br>• Tools and integration with repositories. |
| RQ5 | For which downstream activities are RE work products used? | • Use: specification-driven development, design generation, code generation, test derivation, conformance verification, or reuse of requirements, models, or components.<br>• Traceability: whether it is reported, and between which artifacts. |
| RQ6 | Which requirements quality attributes do the studies consider, and through which mechanisms are they controlled? | • Attribute, following ISO/IEC/IEEE 29148: characteristics of individual requirements (necessary, appropriate, unambiguous, complete, singular, feasible, verifiable, correct, conforming) and of sets of requirements (complete, consistent, feasible, comprehensible, able to be validated), complemented by traceability. Mechanism: human review, automated verification, metrics, testing, or regeneration.<br>• Mechanism: human review, automated verification, metrics, testing, or regeneration. |
| RQ7 | Which research types and validation methods are reported? | • Research type, following Wieringa et al. [21]: solution proposal, validation research, evaluation research, philosophical paper, opinion paper, or experience paper.<br>• Method: experiment, case study, survey, proof of concept, or none.<br>• Setting: academic, industrial, or mixed.<br>• Benefits and limitations: recorded as stated by the authors, without appraisal. |

RQ6 records what the studies report about the quality of the requirements they produce; it does not assess the quality of the studies themselves. Likewise, RQ7 records the reported benefits and limitations as stated by the authors, without appraisal.

In addition, the publication questions in Table 4 characterize the bibliographic and demographic landscape of the field.

Table 4. Publication questions

| PQ# | Publication question | Description |
| --- | --- | --- |
| PQ1 | In which language were the studies published? | To know the distribution of studies by language: English, Spanish or Portuguese. |
| PQ2 | Where have the studies been published? | To know the distribution of studies by type of venue: conference, journal or workshop. |
| PQ3 | How has the quantity of studies evolved over time? | Publications per year. |
| PQ4 | What are the authors' affiliations? | Classify the affiliations into three categories: academy, industry or both. We will consider the affiliations of all the authors. |
| PQ5 | Which countries are the most active? | Considering the author's affiliations (all authors). |

## 4.3. Search strategy

The selected search strategy includes two approaches to look for the primary studies. The first one is an automatic search on the most important online sources for scientific studies (digital libraries and databases). The second one, used in order to ensure the completeness of our set of studies, is the snowballing technique. It comprises the following activities, represented in Figure **1**:

1. **Build control set.** A set of known relevant primary studies is compiled from the related secondary studies (Table 1) and from the researchers' knowledge. It is used to verify that the search strings retrieve the studies they are expected to retrieve.
2. **Select sources.** The digital libraries and complementary sources are confirmed, and access and export formats are verified.
3. **Design the search string.** Search strings (Table **5**) are built from the PICOC (Table 2) and adapted to the syntax of each source.

4. **Run pilot search.** The strings are run with no time restriction to verify the retrieval of the control set and to establish the starting year. If the strings are inadequate, they are redesigned.
5. **Run the searches.** The final strings are run in all sources, with the starting year applied, and the original exports are preserved.

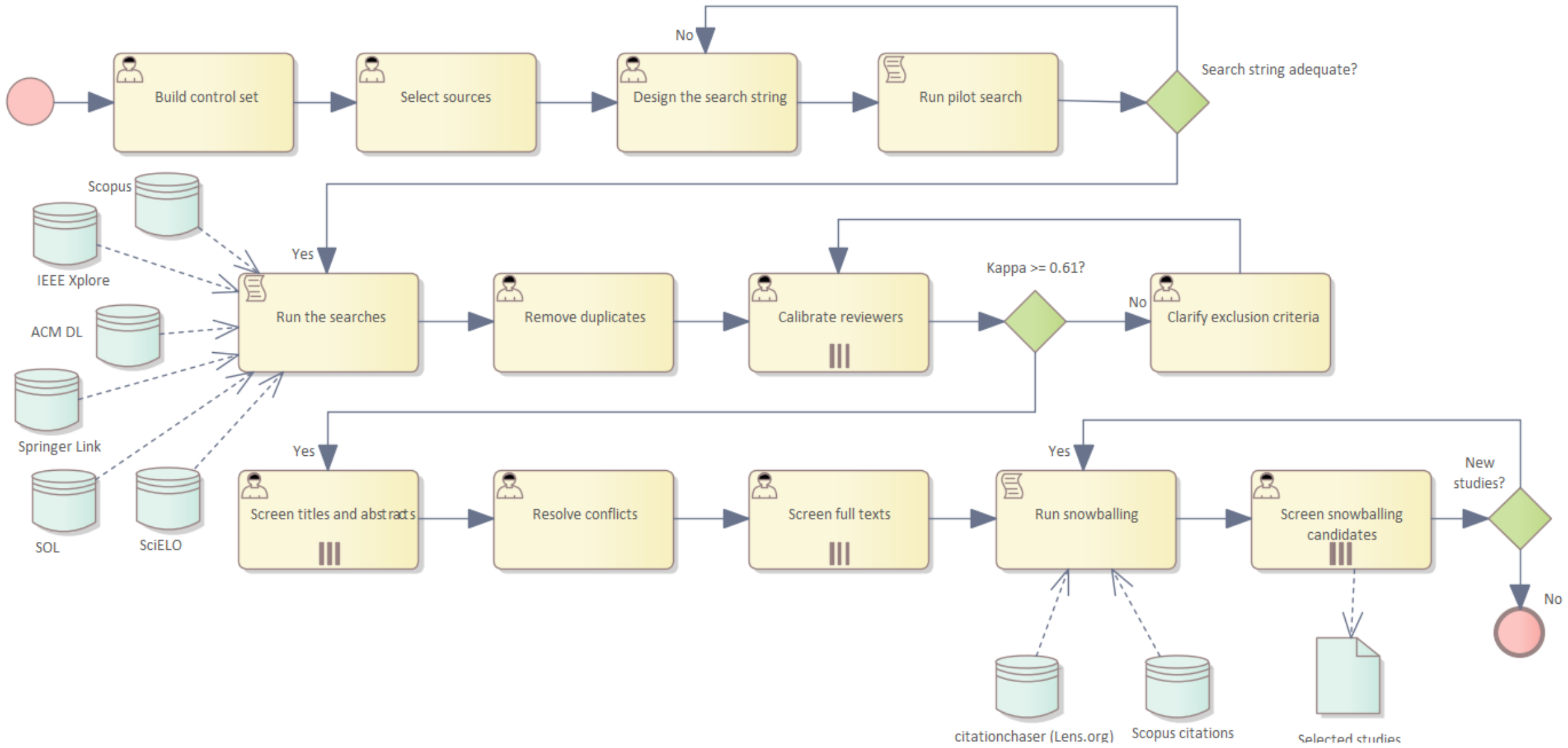

Figure 1. Search and selection process

### 4.3.1. Digital libraries

The automated search is conducted in four digital libraries frequently used in software engineering secondary studies: Scopus, IEEE Xplore, ACM Digital Library, and Springer Link. To capture studies published in Spanish and Portuguese, two complementary sources are also searched: SBC OpenLib (SOL), which hosts the proceedings of the Workshop on Requirements Engineering (WER) and other Brazilian software engineering venues, and SciELO. Search strings are adapted to the syntax of each source and, for the complementary sources, translated into Spanish and Portuguese.

Duplicates are resolved as follows: for works with an extended version, the most recent version is retained; for exact duplicates, the record is retained according to the source priority Scopus, IEEE Xplore, Springer Link, ACM Digital Library, SOL, and SciELO.

### 4.3.2. Search strings

Each string combines two blocks of terms with AND: block A covers RE and its work products, including SDD; block B covers GenAI. The term “use case” is restricted to “use case specification” and “use case model” to avoid noise from expressions such as “use cases of LLMs”. In ACM Digital Library, the search is restricted to title, abstract, and keywords. Table 5 presents the strings, which will be validated and, if necessary, adjusted in the pilot search; any adjustment will be recorded as a protocol amendment.

Table 5. Search strings

| Source | Search string |
|---|---|
| Scopus | TITLE-ABS-KEY("requirements engineering" OR "requirements elicitation" OR "requirements analysis" OR "requirements specification" OR "requirements validation" OR "requirements management" OR "requirements traceability" OR "software requirement*" OR "user stor*" OR "use case specification*" OR "use case model*" OR "acceptance criteria" OR "specification-driven development" OR "spec-driven development") AND TITLE-ABS-KEY("generative AI" OR "generative artificial intelligence" OR "large language model*" OR llm* OR gpt* OR chatgpt OR "foundation model*" OR gemini OR llama OR claude OR copilot OR deepseek OR mistral OR palm) |
| IEEE Xplore | ("requirements engineering" OR "requirements elicitation" OR "requirements analysis" OR "requirements specification" OR "requirements validation" OR "requirements management" OR "requirements traceability" OR "software requirements" OR "user stories" OR "user story" OR "use case specification" OR "use case model" OR "acceptance criteria" OR "specification-driven development" OR "spec-driven development") AND ("generative AI" OR "generative artificial intelligence" OR "large language model" OR "large language models" OR LLM OR LLMs OR GPT OR chatgpt OR "foundation model" OR "foundation models" OR gemini OR llama OR claude OR copilot OR deepseek OR mistral OR palm) |
| ACM DL | (Title:("requirements engineering" OR "requirements elicitation" OR "requirements analysis" OR "requirements specification" OR "requirements validation" OR "requirements management" OR "requirements traceability" OR "software requirements" OR "user stories" OR "user story" OR "use case specification" OR "use case model" OR "acceptance criteria" OR "specification-driven development" OR "spec-driven development") OR Abstract:("requirements engineering" OR "requirements elicitation" OR "requirements analysis" OR "requirements specification" OR "requirements validation" OR "requirements management" OR "requirements traceability" OR "software requirements" OR "user stories" OR "user story" OR "use case specification" OR "use case model" OR "acceptance criteria" OR "specification-driven development" OR "spec-driven development") OR Keyword:("requirements engineering" OR "requirements elicitation" OR "requirements analysis" OR "requirements specification" OR "requirements |

| Source | Search string |
|---|---|
| | validation" OR "requirements management" OR "requirements traceability" OR "software requirements" OR "user stories" OR "user story" OR "use case specification" OR "use case model" OR "acceptance criteria" OR "specification-driven development" OR "spec-driven development")) AND (Title:("generative AI" OR "generative artificial intelligence" OR "large language model" OR "large language models" OR LLM OR LLMs OR GPT OR ChatGPT OR "foundation model" OR "foundation models" OR Gemini OR LLaMA OR Claude OR Copilot OR DeepSeek OR Mistral OR PaLM) OR Abstract:("generative AI" OR "generative artificial intelligence" OR "large language model" OR "large language models" OR LLM OR LLMs OR GPT OR ChatGPT OR "foundation model" OR "foundation models" OR Gemini OR LLaMA OR Claude OR Copilot OR DeepSeek OR Mistral OR PaLM) OR Keyword:("generative AI" OR "generative artificial intelligence" OR "large language model" OR "large language models" OR LLM OR LLMs OR GPT OR ChatGPT OR "foundation model" OR "foundation models" OR Gemini OR LLaMA OR Claude OR Copilot OR DeepSeek OR Mistral OR PaLM)) |
| Springer Link | ("requirements engineering" OR "requirements elicitation" OR "requirements analysis" OR "requirements specification" OR "requirements validation" OR "requirements management" OR "requirements traceability" OR "software requirements" OR "user stories" OR "user story" OR "use case specification" OR "use case model" OR "acceptance criteria" OR "specification-driven development" OR "spec-driven development") AND ("generative AI" OR "generative artificial intelligence" OR "large language model" OR "large language models" OR LLM OR LLMs OR GPT OR ChatGPT OR "foundation model" OR "foundation models" OR Gemini OR LLaMA OR Claude OR Copilot OR DeepSeek OR Mistral OR PaLM) |
| SOL | ("requirements engineering" OR "requirements elicitation" OR "requirements specification" OR "software requirements" OR "user stories" OR "engenharia de requisitos" OR "elicitação de requisitos" OR "especificação de requisitos" OR "requisitos de software" OR "histórias de usuário") AND ("generative AI" OR "large language model" OR "large language models" OR LLM OR ChatGPT OR GPT OR "inteligência artificial generativa" OR "IA generativa" OR "modelos de linguagem") |
| SciELO | ("requirements engineering" OR "requirements elicitation" OR "requirements specification" OR "software requirements" OR "user stories" OR "ingeniería de requisitos" OR "ingeniería de requerimientos" OR "especificación de requisitos" OR "especificación de requerimientos" OR "requisitos de software" OR "requerimientos de software" OR "historias de usuario" OR "engenharia de requisitos" OR "especificação de requisitos" OR "histórias de usuário") AND ("generative AI" OR "large language model" OR "large language models" OR LLM OR ChatGPT OR GPT OR "inteligencia artificial generativa" OR "IA generativa" OR "modelos de lenguaje" OR |

| Source | Search string |
|---|---|
| | "inteligência artificial generativa" OR "modelos de linguagem" OR Gemini OR LLaMA OR Claude OR Copilot OR DeepSeek OR Mistral OR PaLM) |

In Springer Link, which searches full text, the content-type filter (article and conference paper) and the discipline filter (Computer Science) are applied and recorded. For IEEE Xplore, SOL, and SciELO, the pilot search verifies that the strings comply with the term limits and Boolean syntax supported by each source; where a string is not supported, it is split into documented partial searches.

### 4.3.3. Pilot search

A pilot search is run with no time restriction to identify the earliest evidence of the object of this study. That evidence is defined as the earliest study that addresses at least one RE activity using GenAI and is not excluded by criteria EC7 to EC9. The publication year of that study is adopted as the starting year. The pilot search also computes the proportion of the control set retrieved by the strings. The pilot search, the identified study, the control set recall, and the resulting decisions are recorded and reported.

## 4.4. Study selection

### 4.4.1. Exclusion criteria

A study is included if it meets none of the exclusion criteria which are presented in Table 6.

Table 6. Exclusion criteria.

| EC# | Description |
|---|---|
| EC1 | The study is not written in English, Spanish or Portuguese. |
| EC2 | The study is not peer-reviewed. Preprints are accepted only when their acceptance at a peer-reviewed venue can be verified; if the published version is available, it is retained instead of the preprint. |
| EC3 | The study is a short paper (five pages or fewer). |

| EC# | Description |
|---|---|
| EC4 | The study was published before the year of the earliest evidence of GenAI-augmented RE, as established by the pilot search. |
| EC5 | A more recent extended version of the same work exists. |
| EC6 | The study is a secondary or tertiary study. Such studies are considered as related work. |
| EC7 | The study does not address any RE activity. |
| EC8 | GenAI is not part of the proposed approach, its application, or its evaluation. This includes studies that rely exclusively on non-generative AI techniques. |
| EC9 | The study focuses on design, code, or testing without explicitly describing requirements or specifications as a source, constraint, or validation criterion. |

### 4.4.2. Reviewer calibration

Before screening, all reviewers independently assess a common sample of studies. Agreement is measured using Cohen's kappa [22] (two reviewers) or Fleiss' kappa [23] (more than two). Screening begins only when kappa reaches at least 0.61, which corresponds to substantial agreement [24]. Otherwise, the reviewers jointly analyze the disagreements and record written interpretation notes for the criteria, without modifying them, and calibration is repeated with a new sample. Agreement is measured again at the end of title and abstract screening and reported.

### 4.4.3. Screening and conflict resolution

Each study is assessed by at least two reviewers, independently and in blind mode, first on the basis of its title and abstract and then on its full text. Each reviewer classifies the study as include, uncertain, or exclude, indicating the applicable criterion as the exclusion reason. Disagreements are resolved according to the matrix proposed by Petersen et al. [13] (Table 7).

Table 7. Criteria to resolve disagreements.

| | | Researcher 1 | | |
|---|---|---|---|---|
| | | Include | Uncertain | Exclude |
| Researcher 2 | Include | A: included | B: included | D: full reading |
| | Uncertain | B: included | C: full reading | E: excluded |
| | Exclude | D: full reading | E: excluded | F: excluded |

Studies in cases C and D are read in full and reassessed until they reach case A, B, E, or F. The studies included at the end of full-text screening constitute the seed set for snowballing.

### 4.4.4. Snowballing

Backward snowballing (references) and forward snowballing (citations) are applied to the seed set, following Wohlin [17]. Citationchaser [25] is used as the primary tool and Scopus as a complement to improve coverage of software engineering venues. Each seed is processed individually, and the originating seed, direction, iteration, tool, and query date are recorded for each candidate. Candidates are deduplicated against the entire corpus and screened as described in Sections 4.4.1 to 4.4.3. Studies included in an iteration become the seeds of the next one, and the process ends when an iteration yields no new included studies.

## 4.5. Data extraction form

Data are extracted from each primary study into the data extraction form (DEF) presented in Table 8 and Table 9. The values listed are the initial classification schemes; they may be refined through keywording (Section 4.6), and any category added is recorded and applied to all studies. Fields marked as multiple accept more than one value. Each study is extracted by one researcher and verified by a second researcher.

Table 8. Data extraction form for research questions

| Field | RQ | Accepted values |
|---|---|---|
| Study identifier | — | S### (automated search), SB### (snowballing); DOI |
| RE activity (multiple) | RQ1 | Elicitation; analysis; negotiation; specification; validation; management; traceability; evolution; reuse |
| Practice type | RQ1 | Traditional practice augmented by AI; adapted traditional practice; emerging practice |
| Practice description | RQ1 | Free text: objective, activities, inputs, outputs, and context of application |

| Field | RQ | Accepted values |
|---|---|---|
| Process structure | RQ2 | Single task; sequential workflow; iterative workflow; multi-agent system; not reported |
| Degree of automation | RQ2 | Assistance; collaboration; supervised autonomy; full autonomy; not reported |
| Point of human intervention (multiple) | RQ2 | Definition; review; approval; correction; none; not reported |
| Artifact (multiple) | RQ3 | User stories; use cases; functional requirements; non-functional requirements; acceptance criteria; business rules; conceptual or domain models; data dictionaries; structured specifications; other |
| Level of formalization | RQ3 | Informal; semi-structured; structured; formal |
| Notation or format (multiple) | RQ4 | Unconstrained natural language; template-based natural language; UML; BPMN; SysML; formal language; structured format; other |
| Model (multiple) | RQ4 | Model name and version; provider; availability (proprietary or open) |
| Integration approach (multiple) | RQ4 | Zero-shot prompting; few-shot prompting; chain-of-thought prompting; retrieval-augmented generation; fine-tuning; agent-based; other |
| Tools | RQ4 | Free text: tools and integration with repositories |
| Downstream use (multiple) | RQ5 | Specification-driven development; design generation; code generation; test derivation; conformance verification; reuse of requirements; reuse of models; reuse of components; none |
| Traceability | RQ5 | Reported (yes or no); linked artifacts |
| Quality attributes (multiple) | RQ6 | Individual requirement: necessary; appropriate; unambiguous; complete; singular; feasible; verifiable; correct; conforming. Set of requirements: complete; consistent; feasible; comprehensible; able to be validated. Additional: traceable. Other; not reported |
| Control mechanism (multiple) | RQ6 | Human review; automated verification; metrics; testing; regeneration; other; not reported |

| Field | RQ | Accepted values |
|---|---|---|
| Research type | RQ7 | Solution proposal; validation research; evaluation research; philosophical paper; opinion paper; experience paper [21] |
| Empirical method | RQ7 | Experiment; case study; survey; proof of concept; none |
| Setting | RQ7 | Academic; industrial; mixed |
| Reported benefits | RQ7 | Free text, as stated by the authors |
| Reported limitations | RQ7 | Free text, as stated by the authors |
| Notes | — | Free text: extraction remarks and doubts |

Table 9. Data extraction form for publication questions

| Field | PQ | Accepted values |
|---|---|---|
| Language | PQ1 | English; Spanish; Portuguese |
| Venue name | PQ2 | Name of the conference, workshop, or journal |
| Year | PQ3 | Year of publication |
| Affiliation | PQ4 | Academia; industry; both |
| Countries (multiple) | PQ5 | Countries of all authors' affiliations |

## 4.6. Data analysis

The analysis is descriptive, as is appropriate for a mapping study [13]. The initial classification schemes are refined through keywording of abstracts [14]: when the extracted data reveal recurring concepts not covered by the schemes, new categories are proposed, agreed upon by two researchers, recorded, and applied to all studies.

For each dimension, frequencies and percentages are reported. Relationships between dimensions are presented through cross-tabulations and bubble plots, in particular:

- RE activity (RQ1) by downstream use (RQ5);
- artifact type (RQ3) by level of formalization (RQ3) and notation (RQ4);
- integration approach (RQ4) by RE activity (RQ1);
- quality attributes and control mechanisms (RQ6) by artifact type (RQ3);
- research type (RQ7) by RE activity (RQ1) and by year (PQ2).

Publication questions are presented through distributions by venue type, language, year, affiliation, and country. Gaps are identified as combinations of dimensions with few or no studies, particularly between RE work products and their downstream use.

## 4.7. Tool support

Each phase of the study is supported by a designated tool (Table 10). Tool versions and access dates are recorded in the study's traceability dataset.

Table 10. Tool support

| Phase | Tool | Purpose |
|---|---|---|
| Protocol definition | Parsifal | Protocol design and registration. |
| Searches | Scopus, IEEE Xplore, ACM Digital Library, Springer Link, SOL, SciELO | Automated search. |
| Deduplication and screening | Rayyan | Deduplication, independent blind screening, and conflict resolution; exclusion criteria configured as exclusion reasons. |
| Snowballing | Citationchaser, complemented by Scopus | Backward and forward snowballing. |
| Data extraction | Parsifal | Data extraction form. |

| Phase | Tool | Purpose |
|---|---|---|
| Reference management | Zotero | Management of the final set of studies and citation in the report. |
| Traceability | Spreadsheet | Consolidated traceability dataset. |

**Use of generative AI**: GenAI tools may support search string design, translation, data extraction, and analysis. All AI-assisted outputs are verified and adjusted by a researcher. Inclusion and exclusion decisions are made exclusively by researchers. Relevance predictions offered by screening tools are not consulted during independent screening. Every use of AI is logged, recording the tool and model version, date, phase, instruction, output, and verification. The log is published as supplementary material.

# 5. Threats to validity

In order to minimize the impact of the validity threats categorized by Petersen et al. [13] that could affect our study, we present them with the corresponding mitigation actions:

## 5.1. Descriptive validity

Descriptive validity concerns the extent to which observations are described objectively and accurately.

- Data are recorded in a structured data extraction form with predefined accepted values (Table 8 and Table 9).
- An initial meeting is held with all researchers to unify concepts and criteria and to demonstrate the extraction process with examples.
- Each extraction is verified by a second researcher.
- The data extraction form and the extracted data are published.

## 5.2. Theoretical validity

Theoretical validity concerns the ability to capture what is intended to be captured.

- Search coverage. Search strings are tailored to six sources, including two complementary sources for Spanish and Portuguese, and are validated against a control set of known studies during the pilot search.
- Combined search methods. Automated search and iterative backward and forward snowballing are combined to reduce the risk of missing relevant studies [26].

- Terminological variability. GenAI terminology evolves rapidly and SDD has limited academic presence. Block B includes generic and model-specific terms, and SDD is included only as an alternative term within block A.
- Snowballing coverage. Citationchaser relies on Lens.org, whose coverage of software engineering venues may be incomplete; Scopus is used as a complement.
- Selection bias. Exclusion criteria are explicit and operational; reviewers are calibrated with a kappa threshold of 0.61; each study is reviewed independently and in blind mode by at least two researchers; and disagreements are resolved with a predefined matrix.
- Language bias. Studies in English, Spanish, and Portuguese are included; studies in other languages are excluded, which may have a minor impact.
- Publication bias and recency. The field evolves rapidly and many studies first appear as preprints. Only preprints with verifiable peer-reviewed acceptance are included, which may omit the most recent work; this is acknowledged as a limitation.
- Use of GenAI by the researchers. AI support may introduce errors in extraction or translation. All AI-assisted outputs are verified by a researcher, selection decisions are made exclusively by researchers, and every use is logged and reported.

## 5.3. Generalizability

Generalizability concerns the ability to generalize the results to the domain.

- The research questions and classification schemes are general enough to classify studies regardless of application domain, model, or industry [27].
- However, because GenAI models evolve rapidly, results reflect the state of research at the time of the search, which is recorded and reported.

## 5.4. Interpretive validity

Interpretive validity is achieved when conclusions are reasonable given the data.

- Every conclusion is validated by at least two researchers.
- Researchers with experience in RE and in GenAI participate in the interpretation of data.
- New classification categories emerging from keywording are agreed upon by two researchers and applied to all studies.

## 5.5. Repeatability

The research process must be detailed enough to be repeated.

- This protocol is published before the study is conducted.
- Search strings, search dates, filters, and original exports are preserved for each source.
- The traceability dataset records every decision, its rationale, and the reviewers involved.
- Tools, versions, and the log of AI use are reported.
- Any deviation from the protocol is recorded as an amendment (Section 7).

# 6. Scope limitations

Consistent with the nature of a mapping study, the following are outside the scope of this study:

- quality assessment of primary studies;
- synthesis of the effectiveness of practices;
- design of the integrated RE process, which is proposed as future work building on the resulting map.

# 7. Protocol amendments

Deviations from this protocol that arise during the study, such as adjustments to the search strings after the pilot search, the starting year, interpretation notes for the exclusion criteria, or new classification categories, are recorded in an amendment log with their date, description, rationale, and affected phase. The amendment log is published together with the final report.

# 8. Supplementary material

The following material will be made publicly available in an open repository: this protocol and its amendments; search strings, dates, and the number of results per source; the traceability dataset; reviewer agreement values; the data extraction form and extracted data; the snowballing records; and the log of AI use.

# 9. Conclusion

This protocol defines the procedure for a systematic mapping study on generative AI-augmented requirements engineering. Unlike existing secondary studies, which focus on the RE activities supported and the prompting strategies used, this study places the work products of RE, and particularly specifications, at the center of the mapping, and examines how they are used in subsequent activities such as specification-driven development, design and code generation, test derivation, verification, and reuse.

The protocol specifies seven research questions with explicit classification schemes, a search strategy across six sources in three languages validated through a pilot search and a control set, a selection procedure based on operational exclusion criteria, reviewer calibration, and blind independent screening, iterative backward and forward snowballing, a data extraction form, and a descriptive analysis approach. It also defines the tool support, including the controlled use of GenAI by the researchers, and the mitigation of threats to validity.

The resulting map is expected to reveal where research on GenAI-augmented RE is concentrated, which gaps remain between RE work products and their downstream use, and which practices may form the basis of an integrated RE process whose specifications serve as a reliable foundation for the rest of the software life cycle.